\documentclass[aps,nofootinbib,prd,eqsecnum,showpacs,showkeys,preprintnumbers,altaffilletter]{revtex4-1}

\usepackage{amsmath}
\usepackage{amssymb}

\makeatletter

\usepackage{graphicx}
\usepackage{amsfonts}
\usepackage{color}
\usepackage{bm}
\usepackage{mathrsfs}
\usepackage{epstopdf}
\usepackage{url}
\usepackage{footnote}
\usepackage{appendix}

\usepackage{float}
\usepackage{enumerate}
\usepackage[dvipsnames]{xcolor}

\usepackage{hyperref}

\usepackage{booktabs}
\AtBeginDocument{
\heavyrulewidth=.08em
\lightrulewidth=.05em
\cmidrulewidth=.03em
\belowrulesep=.65ex
\belowbottomsep=0pt
\aboverulesep=.4ex
\abovetopsep=0pt
\cmidrulesep=\doublerulesep
\cmidrulekern=.5em
\defaultaddspace=.5em
}

\usepackage{tabularx}

\newcolumntype{L}[1]{>{\hsize=#1\hsize\raggedright\arraybackslash}X}%
\newcolumntype{R}[1]{>{\hsize=#1\hsize\raggedleft\arraybackslash}X}%
\newcolumntype{C}[1]{>{\hsize=#1\hsize\centering\arraybackslash}X}%

\newcommand{\be}{\begin{equation}}
 \newcommand{\ee}{\end{equation}}
\newcommand{\ben}{\begin{eqnarray*}}
 \newcommand{\een}{\end{eqnarray*}}
\newcommand{\bea}{\begin{eqnarray}}
 \newcommand{\eea}{\end{eqnarray}}
\newcommand{\bdm}{\begin{displaymath}}
 \newcommand{\edm}{\end{displaymath}}
\newcommand{\ba}{\begin{align}}
 \newcommand{\ea}{\end{align}}

\makeatother

\begin{document}

\title{Eddington-inspired-Born-Infeld tensorial instabilities neutralized in a quantum approach}

\author{Imanol Albarran $^{1,2}$}
\email{imanol@ubi.pt}

\author{Mariam Bouhmadi-L\'{o}pez $^{1,2,3,4}$}
\email{mariam.bouhmadi@ehu.eus}

\author{Che-Yu Chen $^{5,6}$ }
\email{b97202056@gmail.com}

\author{Pisin Chen $^{5,6,7}$}
\email{pisinchen@phys.ntu.edu.tw}

\date{\today}

\affiliation{
${}^1$Departamento de F\'{\i}sica, Universidade da Beira Interior, Rua Marqu\^{e}s D'\'Avila e Bolama 6200-001 Covilh\~a, Portugal\\
${}^2$Centro de Matem\'atica e Aplica\c{c}\~oes da Universidade da Beira Interior, Rua Marqu\^{e}s D'\'Avila e Bolama 6200-001 Covilh\~a, Portugal\\
${}^3$Department of Theoretical Physics University of the Basque Country UPV/EHU. P.O. Box 644, 48080 Bilbao, Spain\\
${}^4$IKERBASQUE, Basque Foundation for Science, 48011, Bilbao, Spain\\
${}^5$Department of Physics and Center for Theoretical Sciences, National Taiwan University, Taipei, Taiwan 10617\\
${}^6$LeCosPA, National Taiwan University, Taipei, Taiwan 10617\\
${}^7$Kavli Institute for Particle Astrophysics and Cosmology, SLAC National Accelerator Laboratory, Stanford University, Stanford, CA 94305, USA\\
}

\begin{abstract}
The recent direct detection of gravitational waves has highlighted the huge importance of the tensorial modes in any extended gravitational theory. One of the most appealing approaches to extend gravity beyond general relativity is the Eddington-inspired-Born-Infeld gravity which is formulated within the Palatini approach. This theory can avoid the big bang singularity in the physical metric although a big bang intrinsic to the affine connection is still there, which in addition couples to the tensorial sector and might jeopardize the viability of the model. In this paper, we suggest that a quantum treatment of the affine connection, or equivalently of its compatible metric, is able to rescue the model. We carry out such an analysis and conclude that from a quantum point of view such a big bang is unharmful. We expect therefore that the induced tensorial instability, caused by the big bang intrinsic to the affine connection, can be neutralized at the quantum level.
\end{abstract}



\maketitle

\section{Introduction}
It is commonly recognized that Einstein's general relativity (GR), though very successful in describing our universe, nonetheless suffers from several fundamental puzzles. On the very early stage of the universe where the energy scale and the curvature scale were huge, say, close to the Planck scale, a purely classical description of gravity based on GR would not be sufficient. Actually, it is expected that a fundamental quantum theory of gravity is necessary such that some pathologies of GR at high energy scales can be resolved, such as the non-renormalizability of the theory and the issue regarding spacetime singularities. Whereas, it is still not clear so far how such a fundamental quantum gravity theory should be built in a self-consistent way. The development of a complete quantum theory of gravity is still an open question and it is certainly one of the most active research directions in modern theoretical physics.   

From a more conservative point of view, to escape from the aforementioned theoretical swamp, one may resort to other modified theories beyond GR and regard them as effective theories of the unknown quantum theory of gravity \cite{Capozziello:2011et}. It is likely that such extended theories of gravity, even presumably not complete, are already able to ameliorate the UV problems in GR. Among the plethora of extended theories of gravity, the Eddington-inspired-Born-Infeld gravity (EiBI) proposed in \cite{Banados:2010ix} is appealing in several theoretical aspects. First, it reduces exactly to GR in vacuum and deviates from it when matter fields are included. Second, due to the square root structure in the gravitational action, the curvature scale and the energy scale seem to be bounded from above and the big bang singularity is naturally avoided in the EiBI gravity. Third, the theory is simple in the sense that it only contains one free additional parameter, the Born-Infeld constant $\kappa$ compared with GR. Fourth, it is free of ghost instabilities because the theory is constructed through the Palatini rather than the metric variational principle. Actually, the idea of including the Born-Infeld structure into the gravitational theory was proposed in Ref.~\cite{Deser:1998rj}. However, the theory is built with the metric variational principle and it has ghost because of the higher order derivative terms in the field equations. The EiBI theory, on the contrary, is formulated via the Palatini variational principle. The field equations only contain second order derivatives and therefore no ghost is present in the theory. The applications and several properties of the EiBI gravity have been studied widely in the literature \cite{Casanellas:2011kf,Avelino:2012ge,Delsate:2012ky,Liu:2012rc,EscamillaRivera:2012vz,Avelino:2012ue,Avelino:2012qe,Cho:2012vg,Scargill:2012kg,Bouhmadi-Lopez:2013lha,Cho:2013pea,Harko:2013wka,Yang:2013hsa,Olmo:2013gqa,Sham:2013cya,Du:2014jka,Makarenko:2014lxa,Fernandes:2014bka,Wei:2014dka,Odintsov:2014yaa,Bouhmadi-Lopez:2014jfa,Jimenez:2014fla,Sotani:2014lua,Chen:2015eha,Sotani:2015ewa,Li:2017ttl,BeltranJimenez:2017uwv,Latorre:2017uve,Bouhmadi-Lopez:2017lbx,Chen:2017ify,Jana:2017ost,Roshan:2018pts,Chen:2018mkf,Shaikh:2018yku,Chen:2018vuw,Afonso:2018mxn}. Some attempts to quantize the EiBI gravity have been proposed in Refs.~\cite{Bouhmadi-Lopez:2016dcf,Arroja:2016ffm,Albarran:2017swy,Bouhmadi-Lopez:2018tel,Bouhmadi-Lopez:2018sto,Albarran:2018mpg}. See also \cite{BeltranJimenez:2017doy} for a nice review on the EiBI gravity and other interesting Born-Infeld inspired theories of gravity. A further motivation to consider the EiBI theory is that the Born-Infeld type of theories, of which EiBI gravity is a subclass, have intrinsic Noether symmetries as shown in \cite{Capozziello:2014qla}. This is not surprising as the same happens in other modified theories of gravity \cite{Bouhmadi-Lopez:2016fgd}. For an interesting review on Noether symmetry, please see \cite{Capozziello:1996bi}.

In this paper, we will highlight an important issue regarding the viability of the EiBI theory. In the EiBI gravity, the big bang singularity in the physical metric is avoided by hiding the divergences of quantities in the second spacetime structure defined by the affine connection. The physical metric $g_{\mu\nu}$ is non-singular while the other metric, which we will call the auxiliary metric $q_{\mu\nu}$ later, turns out to be singular. Since the matter field is assumed to be coupled only to the physical metric, the singularity in the auxiliary metric seems to be unharmful for a physical observer. However, it has been proven in Refs.~\cite{Yang:2013hsa,EscamillaRivera:2012vz} that the metric perturbations, especially the tensor perturbations, are actually unstable for the non-singular solutions in the EiBI gravity, jeopardizing the validity of the theory. A more careful analysis in Ref.~\cite{BeltranJimenez:2017uwv} reveals that the propagation of gravitational waves does see the structure of the auxiliary metric and it is the singularity in the auxiliary metric that gives rise to the linear instabilities of the theory. It should be noted that the problem of tensor instabilities mentioned above can be ameliorated for a positive Born-Infeld coupling constant if a time-dependent equation of state parameter is considered \cite{Avelino:2012ue}.

In order to resolve this problem, we will suggest a quantum treatment to the EiBI gravity in the framework of quantum geometrodynamics. In this approach, the construction of the Wheeler DeWitt (WDW) equation is crucial since the WDW equation describes the quantum evolution of the universe as a whole \cite{KieferQG}. The derivation of the WDW equation stems from a self-consistent classical Hamiltonian, which and all the phase space functions are then promoted to quantum operators. The Hamiltonian, being a first class constraint of the system, turns out to be a restriction on the Hilbert space which is exactly the WDW equation.  The strategy is to see whether the wave function would vanish near the configuration of the singularity in the auxiliary metric, satisfying the DeWitt (DW) boundary condition \cite{DeWitt:1967yk}. If the answer is yes, it is then expected that the singularity can be avoided in the quantum world and the linear instabilities, which result from this singularity, can be naturally resolved. For the sake of completeness, we will consider two kinds of matter descriptions, one is the perfect fluid description and the other is the scalar field description. For the perfect fluid description, the matter field is governed by a perfect fluid with a constant equation of state. The system contains only one degree of freedom corresponding to the scale factor of the metric. As for the scalar field description, a scalar field degree of freedom is included into the system and the WDW equation turns out to be a partial differential equation with two independent variables. For each description and each non-singular solution, we will solve the corresponding WDW equation and we will exhibit that the DW condition can always be satisfied, indicating the resolution of the singularity in the auxiliary metric as well as the instabilities via quantum effects.

This paper is outlined as follows. In section \ref{class.sec}, we briefly review the non-singular cosmological solutions in the EiBI gravity, depending on different signs of the Born-Infeld parameter $\kappa$. We will also exhibit how the tensor instabilities are related to the singularity of the auxiliary metric. In section \ref{section3p}, we use the perfect fluid description and derive the WDW equations of the non-singular solutions with regard the physical metric for each sign of $\kappa$. The WDW equations within the scalar field description are obtained in section \ref{section4s}. After deriving the WDW equation, we will obtain the wave function and see whether the solution satisfies the DW boundary condition near the singularity of the auxiliary metric. For the perfect fluid description and the scalar field description, the WDW equations will be solved, respectively, in sections \ref{solvwdwp} and \ref{solvescalar}. We finally conclude in section \ref{conclusionsec}.

\section{The classical universe: big bang in the auxiliary metric}\label{class.sec}
The EiBI gravity is formulated by the following action \cite{Banados:2010ix}:
\begin{equation}
\mathcal{S}_{\textrm{EiBI}}=\frac{2}{\kappa}\int d^4x\left[\sqrt{|g_{\mu\nu}+\kappa R_{(\mu\nu)}|}-\lambda\sqrt{-g}\right]+\mathcal{S}_M\left(g,\Psi\right)\,,\label{eibiaction}
\end{equation}
where $S_M$ is the matter Lagrangian of the matter field $\Psi$ and it is assumed to be coupled only to the physical metric $g_{\mu\nu}$. There is only the symmetric part of the Ricci curvature tensor $R_{(\mu\nu)}$ appearing in the action and the curvature is constructed by the affine connection $\Gamma$, which is in principle independent of $g_{\mu\nu}$. Based on these assumptions, the theory respects the projective symmetry and one does not have to consider torsion fields since they can be removed by imposing a gauge fixing condition \cite{BeltranJimenez:2017doy}. Moreover, in the action \eqref{eibiaction}, $|g_{\mu\nu}+\kappa R_{(\mu\nu)}|$ is the absolute value of the determinant of the tensor $g_{\mu\nu}+\kappa R_{(\mu\nu)}$, where $\kappa$ characterizes the theory and has inverse dimensions to that of a cosmological constant.{\footnote{In this paper, we assume $8\pi G=c=1$.}} Finally, the dimensionless constant $\lambda$ qnantifies the effective cosmological constant at the low curvature limit.

Since the theory is formulated within the Palatini variational principle, one has to vary the action with respect to the physical metric and the affine connection separately. It turns out that one can define an auxiliary metric satisfying $\lambda q_{\mu\nu}=g_{\mu\nu}+\kappa R_{(\mu\nu)}$ such that $q_{\mu\nu}$ is compatible with the affine connection. One of the field equations relates algebraically the matter field to the two metrics, and the other equation is a second order differential equation of $q_{\mu\nu}$. It can be seen that when the curvature vanishes, the two metrics are identical up to a \textit{constant} conformal rescaling, rendering the equivalence of the EiBI theory and GR in the zero curvature regime.  

It is well known that the EiBI gravity reduces to Einstein GR when matter fields are absent. However, the theory could have significant differences from GR when, say, the curvature and the energy density of the matter field take large values. Essentially, that is how the big bang singularity is removed in the EiBI gravity. The existence of the affine structure and its corresponding auxiliary metric actually plays a crucial role in the avoidance of singularities. The divergences of the physical metric at the singularity are \textit{transferred} to the auxiliary spacetime, leaving the physical metric $g_{\mu\nu}$ non-singular. Since the matter field only sees the spacetime structure of the physical metric, the \textit{hidden} singularity in the auxiliary metric seems unharmful for physical observers. Depending on the sign of the parameter $\kappa$, the big bang singularity can be replaced with a bouncing solution in the physical metric when $\kappa<0$, or can be replaced with a \textit{loitering stage} in which the universe acquires its minimum size in the asymptotic past when $\kappa>0$. Table~\ref{summary} briefly summarizes how the EiBI gravity cures the big bang singularity in the physical metric, and also points out the singularity appearing in the auxiliary metric.

\begin{table*}
 \begin{center}
  \begin{tabular}{||c|c|c||}
  \hline 
   $\kappa$ & Physical metric $g_{\mu\nu}$ & Auxiliary metric $q_{\mu\nu}$\\
  \hline\hline 
   Positive & Loitering effects & Big Bang singularity\\ 
  \hline
   Negative & Bounce & Big Bang singularity \\
  \hline
  \end{tabular}
  \caption{This table summarizes how the EiBI theory of gravity cures the big bang singularity in a radiation dominated universe. If $\kappa>0$, the big bang singularity in the physical metric is replaced with a loitering stage in which the universe gets its minimum size in the infinite past. If $\kappa<0$, the physical metric is described by a bouncing scenario in the past. However, there is still a big bang singularity in the auxiliary metric for both cases.}
    \label{summary}
 \end{center}
\end{table*}

In the following subsections, we will briefly review how the big bang singularity is avoided in the EiBI gravity with different signs of $\kappa$, and we shall point out the fact that the big bang singularity actually migrates to the \textit{auxiliary} spacetime, i.e., the curvature invariants defined by the auxiliary metric diverge and the scale factor of the auxiliary metric is zero. We will illustrate it by considering a homogeneous and isotropic universe filled with a perfect fluid with a constant and positive equation of state $w>0$. Then, we will mention how the instability issues in the physical metric arise alongside the auxiliary singularity, which motivates us for the quantum analysis in this paper.

\subsection{The big bang singularity in the auxiliary metric with $\kappa<0$}\label{clnegativekappa}
We first consider a homogeneous and isotropic universe which can be described by the following metric ansatz:
\begin{equation}
ds_g^2=-N(t)^2dt^2+a(t)^2\delta_{ij}dx^idx^j\,,\qquad
ds_q^2=-N(t)^2dt^2+a(t)^2\delta_{ij}dx^idx^j\,,
\end{equation} 
where $N(t)$ and $a(t)$ are the lapse function and the scale factor of the physical metric $g_{\mu\nu}$, while $M(t)$ and $b(t)$ represent the lapse function and the scale factor of the auxiliary metric $q_{\mu\nu}$. These functions are functions of the cosmic time $t$ and they can be determined by the Euler-Lagrange equations of motion. For later convenience, we will define two new variables
\begin{equation}
X\equiv\frac{N}{M}\,,\qquad Y\equiv\frac{a}{b}\,.
\end{equation}

If the universe is dominated by a perfect fluid with energy density $\rho$ and pressure $p=w\rho$, the equations of motion obtained from varying the action with respect to $g_{\mu\nu}$ relate algebraically the two metrics to the matter sector as follows
\begin{equation}
\frac{\lambda X}{Y^3}=\lambda+\kappa\rho\,,\qquad\frac{\lambda}{XY}=\lambda-\kappa p=\lambda-\kappa w\rho\,.\label{classicalconstr}
\end{equation}
From the above equations \eqref{classicalconstr}, one can see that when $\kappa<0$ the energy density of the perfect fluid $\rho$ is bounded from above by
\begin{equation}
\lambda+\kappa\rho\ge0\,,\qquad\Longrightarrow\qquad |\kappa|\rho\le\lambda\,.
\end{equation}
Since the energy density $\rho\propto a^{-3(1+w)}$ is bounded from above, the physical scale factor has a minimum value $a_{m1}$ satisfying $\rho\left(a_{m1}\right)=\lambda/|\kappa|$. The Hubble rate defined by the physical metric $H\equiv\dot{a}/a$, where the dot denotes the derivative with respect to $t$, reads \cite{Scargill:2012kg}
\begin{equation}
H^2\approx\frac{8N^2\left(a-a_{m1}\right)}{3|\kappa|a_{m1}}\,,\label{bouncehubble}
\end{equation}
when $a\rightarrow a_{m1}$. By assuming a constant lapse function $N$, it can be proven that the big bang singularity in the physical metric is replaced with a bouncing solution in the sense that Eq.~\eqref{bouncehubble} can be integrated to get $a-a_{m1}\propto t^2$.

To study the behavior of the auxiliary metric when $a\rightarrow a_{m1}$, we rewrite Eqs.~\eqref{classicalconstr} as follows
\begin{equation}
\frac{\lambda^2}{Y^4}=\left(\lambda+\kappa\rho\right)\left(\lambda-\kappa w\rho\right)\,,\qquad
\frac{X^4}{\lambda^2}=\frac{\lambda+\kappa\rho}{\left(\lambda-\kappa w\rho\right)^3}\,.\label{5}
\end{equation}
When $\rho\rightarrow\lambda/|\kappa|$ and $a\rightarrow a_{m1}$, Eqs.~\eqref{5} can be written as
\begin{equation}
b=\frac{a}{Y}\approx\frac{a_{m1}\left(1+w\right)^{1/4}}{\lambda^{1/4}}\left(\lambda+\kappa\rho\right)^{1/4}\rightarrow0\,,\qquad
X=\frac{N}{M}\approx\frac{b}{a_{m1}\left(1+w\right)}\rightarrow 0\,.\label{8}
\end{equation}
Therefore, at the bounce where $a=a_{m1}$, the auxiliary scale factor $b$ vanishes. On the other hand, the scalar curvature defined by the auxiliary metric is given by
\begin{align}
R[q]\equiv q^{\mu\nu}R_{(\mu\nu)}=\frac{1}{\kappa}\left(4\lambda-X^2-3Y^2\right)\,.
\end{align}
When $a=a_{m1}$, it can be seen that $R[q]$ diverges because $Y\rightarrow\infty$. Also, by suitably choosing the lapse functions, it can be shown that this divergence happens at a finite time $t$. Therefore, there is a big bang singularity in the auxiliary metric when $b=0$. 

\subsection{The big bang singularity in the auxiliary metric with $\kappa>0$}
If $\kappa$ is positive, the big bang singularity in the physical metric is again avoided in the EiBI gravity but in a different manner. In this case, according to Eqs.~\eqref{classicalconstr}, the energy density of the perfect fluid $\rho$ is bounded from above by
\begin{equation}
\lambda-\kappa w\rho\ge0\,,\qquad\Longrightarrow\qquad \kappa\rho\le\frac{\lambda}{w}\,.
\end{equation}
Therefore, the physical scale factor has a minimum value $a_{m2}$ satisfying $\rho\left(a_{m2}\right)=\lambda/\left(w\kappa\right)$. The Hubble rate defined by the physical metric can be approximated as \cite{Scargill:2012kg}
\begin{equation}
H^2\approx\frac{8N^2\left(a-a_{m2}\right)^2}{3\kappa a_{m2}^2}\,,\label{loiteringhubble}
\end{equation}
when $a\rightarrow a_{m2}$. By assuming a constant lapse function $N$, Eq.~\eqref{loiteringhubble} can be integrated to get $a-a_{m2}\propto e^t$. It can be seen that   the scale factor $a$ takes its minimum value when $t\rightarrow-\infty$. The big bang singularity in the physical metric is thus avoided. 

Regarding the asymptotic behavior of the auxiliary metric when $a\rightarrow a_{m2}$, we use Eqs.~\eqref{5} and consider the limit where $\rho\rightarrow\lambda/\left(w\kappa\right)$ to get
\begin{equation}
b\approx\frac{a_{m2}\left(1+w\right)^{1/4}}{\left(w\lambda\right)^{1/4}}\left(\lambda-\kappa w\rho\right)^{1/4}\rightarrow0\,,\qquad
X\approx\frac{\left(1+w\right)a_{m2}^3}{wb^3}\rightarrow \infty\,.\label{12}
\end{equation}
Therefore, the auxiliary scale factor $b$ vanishes and it can be shown that the auxiliary curvature diverges when $b\rightarrow 0$. Also, by suitably choosing the lapse functions, it can be proven that this divergence happens at a finite time $t$. Therefore, there is a big bang singularity in the auxiliary metric.

\subsection{The instability of linear perturbations}
In the EiBI gravity, the big bang singularity in the physical metric can be avoided in the sense that the matter field is minimally coupled with the physical metric, hence the physical observers can only see the geometry of that metric, which is free of the big bang singularity. However, non-singular behaviors of the physical metric in the EiBI gravity are still problematic because of the tensor instabilities. Actually, it has been proven in Ref.~\cite{BeltranJimenez:2017uwv} that these tensor instabilities are highly related to the singular behaviors of the auxiliary metric. In other words, the propagation of gravitational waves would be affected by the geometry of the auxiliary metric. The tensor instabilities in the EiBI gravity were firstly found in Ref.~\cite{EscamillaRivera:2012vz}. In addition, the instabilities of scalar mode and vector mode perturbations have been discovered in Ref.~\cite{Yang:2013hsa}. In this subsection, we will briefly review the tensor instabilities of the non-singular solutions in the EiBI gravity and it will become clear that these instabilities are indeed related to the singularity in the auxiliary metric.  

Considering the tensor perturbations of the metrics such that the perturbed metrics are $\delta g_{ij}=a^2h_{ij}$ and $\delta q_{ij}=b^2\gamma_{ij}$, it has been proven in Ref.~\cite{Jimenez:2015caa} that in the absence of any anisotropic stress, the transverse-traceless tensor perturbations of the two metrics are equivalent, that is, $h_{ij}=\gamma_{ij}$. The evolution of the tensor perturbation is described by the following equation \cite{EscamillaRivera:2012vz,Yang:2013hsa,BeltranJimenez:2017uwv,Bouhmadi-Lopez:2017lbx}:
\begin{align}
&\ddot{h}_{ij}+\left(\frac{3\dot{b}}{b}-\frac{\dot{M}}{M}\right)\dot{h}_{ij}-\frac{M^2}{b^2}\nabla^2h_{ij}\nonumber\\
=&\,\ddot{h}_{ij}+\left(3H-\frac{3\dot{Y}}{Y}-\frac{\dot{N}}{N}+\frac{\dot{X}}{X}\right)\dot{h}_{ij}-\frac{N^2Y^2}{X^2a^2}\nabla^2h_{ij}=0\,.\label{213tensor}
\end{align}
It can be seen that the propagation of the tensor mode is able to see the geometry of the auxiliary metric. When $\kappa<0$, by using Eqs.~\eqref{bouncehubble} and \eqref{8}, it can be proven that the coefficient of the friction term in Eq.~\eqref{213tensor}, that is, the $3\dot{b}/b-\dot{M}/M$ term, is proportional to $1/b^2$ when $b\rightarrow0$. The coefficient of the last term, i.e., $M^2/b^2$, is proportional to $1/b^4$ in the same limit. Therefore, these two terms diverge and the tensor perturbation is severely unstable at the physical bounce. On the other hand, for a positive $\kappa$, the coefficients of both the friction term and the last term are proportional to $b^4$ when $b\rightarrow0$. In this regard, the tensor mode behaves linearly in time. Therefore, when approaching the loitering stage at $t\rightarrow-\infty$, the tensor mode grows linearly backward in time and it diverges in the asymptotic past. 

As can be seen above and according to the results in Ref.~\cite{BeltranJimenez:2017uwv}, the instabilities of the tensor perturbations in the EiBI gravity indeed result from the divergence of the auxiliary metric. Even though the physical observers can only see the non-singular metric, the linear instabilities still jeopardize the validity of the theory. This motivates us to study whether the hidden singularity in the auxiliary metric can be resolved by including some sorts of quantum effects and we will address this issue in the following sections.  

\section{The WDW equation: perfect fluid}\label{section3p}
As mentioned in the previous section, the instability of linear perturbations in the EiBI gravity is highly associated with the divergences appearing in the auxiliary metric. Therefore, as long as such a singularity can be ameliorated by quantum effects, the instability problems can be naturally resolved. To address this issue, we shall consider a quantum geometricodynamical approach in which the WDW equation plays a central role. To derive the WDW equation, one is supposed to start with the correct classical Hamiltonian $\mathcal{H}_T$, which gives the classical equations of motion, and then promote the Hamiltonian to a quantum operator: $\mathcal{H}_T\rightarrow\mathcal{\hat{H}}_T$. In this regard, it can be proven that the Hamiltonian stands for a first class constraint, indicating that it corresponds to a restriction on the Hilbert space, more precisely, $\mathcal{\hat{H}}_T|\psi\rangle=0$.

In Refs.~\cite{Delsate:2012ky,BeltranJimenez:2017doy}, it was shown that the EiBI action \eqref{eibiaction} can be transformed to its Einstein frame via a Legendre transformation. After such a transformation, the new action reads
\begin{equation}
\mathcal{S}_a=\lambda\int d^4x\sqrt{-q}\left[R[q]-\frac{2\lambda}{\kappa}+\frac{1}{\kappa}\left(q^{\alpha\beta}g_{\alpha\beta}-2\sqrt{\frac{g}{q}}\right)\right]+\mathcal{S}_M\left(g,\Psi\right)\,.\label{alteraction}
\end{equation}
On the above action \eqref{alteraction}, the fundamental variables are $g_{\mu\nu}$ and the auxiliary metric $q_{\mu\nu}$. It can be proven that the equations of motion derived from the original action \eqref{eibiaction} can be obtained unambiguously by varying the action \eqref{alteraction} with respect to $g_{\mu\nu}$ and $q_{\mu\nu}$. In our previous papers \cite{Bouhmadi-Lopez:2016dcf,Bouhmadi-Lopez:2018tel,Albarran:2017swy,Albarran:2018mpg}, we have used this alternative action \eqref{alteraction} to deduce the classical Hamiltonian and the corresponding WDW equation in the EiBI gravity. It turns out that the construction of the WDW equation is much more straightforward because of the absence of the square root structure of the curvature present in the original action \eqref{eibiaction}.

Using the alternative action \eqref{alteraction} and assuming that the matter sector is described by a perfect fluid with energy density $\rho$ and pressure $p$, the reduced Lagrangian associated with the action \eqref{alteraction} can be written as
\begin{equation*}
\mathcal{L}=\lambda Mb^3\left[-\frac{6\dot{b}^2}{M^2b^2}-\frac{2\lambda}{\kappa}+\frac{1}{\kappa}\left(X^2+3Y^2-2XY^3\right)\right]-2\rho\left[\left(bY\right)\right]Mb^3XY^3\,.
\end{equation*}
Note that the energy density is expressed as a function of the physical scale factor $a$ and the relation $a=bY$ has been imposed. According to the definition of conjugate momenta, we have three primary constraints:
\begin{equation}
p_X=\frac{\partial\mathcal{L}}{\partial\dot{X}}\sim0\,,\qquad
p_Y=\frac{\partial\mathcal{L}}{\partial\dot{Y}}\sim0\,,\qquad
p_M=\frac{\partial\mathcal{L}}{\partial\dot{M}}\sim0\,,
\end{equation}
where $\sim$ denotes the weak equality, i.e., the equality on the constraint surface. The total Hamiltonian is defined as follows
\begin{align}
\mathcal{H}_T=&\,M\left[-\frac{p_b^2}{24\lambda b}+\frac{2\lambda^2b^3}{\kappa}-\frac{\lambda}{\kappa}b^3X^2-\frac{3\lambda}{\kappa}b^3Y^2+\frac{2XY^3b^3}{\kappa}\left(\lambda+\kappa\rho\right)\right]\nonumber\\
&+\lambda_Xp_X+\lambda_Yp_Y+\lambda_Mp_M,
\end{align}
where $p_b$ is the conjugate momentum of the phase space variable $b$. In the last few terms, $\lambda_X$, $\lambda_Y$, and $\lambda_M$ are Lagrange multipliers associated with each primary constraint.  

In Refs.~\cite{Bouhmadi-Lopez:2018tel,Albarran:2017swy}, a thorough constraint analysis of this system has been carried out. In Ref.~\cite{Albarran:2018mpg}, an improved investigation has been done in which the matter sector is assumed to be a scalar field rather than a perfect fluid. As expected, the Hamiltonian itself is a first class constraint and at the quantum level, it would be treated as a restriction on the Hilbert space, giving rise to the WDW equation. In addition, the equations of motion \eqref{classicalconstr}, which relate algebraically the metrics and the energy-momentum tensor, are exactly the secondary constraints of the system and they are second class constraints. In the presence of second class constraints, one has to resort to the Dirac brackets to promote the phase space functions to quantum operators \cite{Diraclecture}. By doing so, the second class constraints can be directly regarded as zero operators and the WDW equation can be significantly simplified.

As a result, we can find a basis $\langle b|$ to write down the WDW equation $\langle b|\hat{\mathcal{H}}_T|\psi\rangle=0$ as follows
\begin{equation}
\frac{-1}{24\lambda}\langle b|\frac{\hat{p}_b^2}{b}|\psi\rangle+V(b)\langle b|\psi\rangle=0\,.
\label{WDWgeneral}
\end{equation}
Note that the variables $X$ and $Y$ can be expressed as functions of $b$ by imposing the second class constraints given by Eqs.~\eqref{classicalconstr} in this model (see also Eqs.~(3.9) and (3.10) in Ref.~\cite{Bouhmadi-Lopez:2018tel}). Therefore, the potential $V(b)$ can be expressed as follows
\begin{equation}
V(b)=\frac{2\lambda^2b^3}{\kappa}+\frac{\lambda}{\kappa}b^3X^2(b)-\frac{3\lambda}{\kappa}b^3Y^2(b)\,.\label{wdwpotential}
\end{equation}

\subsection{The WDW equation for $\kappa<0$}
In the perfect fluid description, Eq.~\eqref{WDWgeneral} and the potential \eqref{wdwpotential} stands for a general expression of the WDW equation of the EiBI gravity. The explicit expression of $X(b)$ and $Y(b)$ are given by the constraints \eqref{classicalconstr} and they depend on the cosmological solutions under consideration. In this subsection, we focus on the approximated cosmological solutions of the bouncing scenario for a negative $\kappa$, which has been discussed in subsection \ref{clnegativekappa}. In this case, we insert the approximated behaviors \eqref{8} to the potential $V(b)$ and the potential can be approximated as 
\begin{equation}
V(b)=-\frac{2\lambda^2b^3}{|\kappa|}-\frac{\lambda b^5}{|\kappa|a_{m1}^2\left(1+w\right)^2}+\frac{3\lambda a_{m1}^2 b}{|\kappa|}\,,\label{17}
\end{equation}
when $b\rightarrow0$. It can be also seen that when $b\rightarrow0$, the last term on the right hand side dominates.

We choose the following factor ordering
\begin{equation}
\frac{\hat{p}_b^2}{b}=-\hbar^2\left(\frac{1}{\sqrt{b}}\frac{\partial}{\partial b}\right)\left(\frac{1}{\sqrt{b}}\frac{\partial}{\partial b}\right)\label{factorordering}
\end{equation}
and introduce a new variable
\begin{equation}
y=\left(c_1b\right)^{3/2}\,,\qquad\textrm{where}\qquad c_1^4=\frac{a_{m1}^2\lambda^2}{|\kappa|\hbar^2}\,.
\end{equation}
The WDW equation when $b\rightarrow0$ ($y\rightarrow0$) can be written as
\begin{equation}
\left(\frac{d^2}{dy^2}+32y^{\frac{2}{3}}\right)\psi(y)=0\,.\label{wdwperfctflkn}
\end{equation}
Note that only the last term on the right hand side of Eq.~\eqref{17} is considered because it dominates the potential when $b\rightarrow0$.

\subsection{The WDW equation for $\kappa>0$}
In this subsection, we shall derive the approximated WDW equation when $b\rightarrow0$ in the EiBI gravity with a positive $\kappa$. In this case, we insert the asymptotic equations \eqref{12} into the potential \eqref{wdwpotential}, and the potential can be approximated as
\begin{equation}
V(b)=\frac{2\lambda^2b^3}{\kappa}+\frac{\lambda\left(1+w\right)^2a_{m2}^6}{\kappa w^2 b^3}-\frac{3\lambda a_{m2}^2 b}{\kappa}\,.\label{22}
\end{equation}
It can be seen that when $b\rightarrow0$, the second term on the right hand side dominates.

To proceed, we choose the same factor ordering as Eq.~\eqref{factorordering}, and introduce a new variable
\begin{equation}
z=b^{3/2}\,.
\end{equation}
The WDW equation when $b\rightarrow0$ ($z\rightarrow0$) can be written as
\begin{equation}
\left(\frac{d^2}{dz^2}+\frac{c_2}{z^2}\right)\psi(z)=0\,,\label{wdwpefkp}
\end{equation}
where $c_2$ is a positive constant and it is defined as
\begin{equation}
c_2=\frac{32\lambda^2\left(1+w\right)^2a_{m2}^6}{3\kappa\hbar^2w^2}\,.\label{c2definition}
\end{equation}
Note that only the second term on the right hand side of Eq.~\eqref{22} is considered since it dominates the potential when $b\rightarrow0$.

\section{The WDW equation: scalar field}\label{section4s}
In the previous section, we have derived the WDW equations near the singularity of the auxiliary metric by using a perfect fluid description. The quantum system has only one degree of freedom in the sense that the WDW equation turns out to be an ordinary differential equation of a single variable, the scale factor $b$. However, the assumption of the perfect fluid description is just for convenience and may not be complete to describe the quantum evolution of the universe in a satisfactory manner. For the sake of completeness, in this section we will introduce an additional degree of freedom, the scalar field $\phi$ minimally coupled to the EiBI gravity, to describe the matter sector of the gravitational system. In the classical regime, it is well-known that the properties of a perfect fluid, including its equation of state and evolution, can be described by a scalar field when a corresponding potential $V(\phi)$ is chosen. In the quantum regime, on the other hand, the two degrees of freedom from the geometrical sector and from the matter sector do not necessarily relate to each other as in the classical regime. In this regard, the WDW equation becomes a partial differential equation of two variables $b$ and $\phi$. To have a more complete picture of the quantum behavior of the universe near the singularity, we will solve the wave function and investigate how the wave function evolves in the two dimensional $(b,\phi)$ space. We shall mention that in Ref.~\cite{Albarran:2018mpg}, we have studied the quantum avoidance of the big rip singularity in the EiBI gravity by solving the WDW equation with two degrees of freedom, one from the geometrical sector and the other is the phantom scalar field from the matter sector.
 
Considering a homogeneous and isotropic metric and assuming a scalar field minimally coupled to the gravity sector, the reduced Lagrangian can be written as follows \cite{Albarran:2018mpg}:
\begin{equation}
\mathcal{L}=\lambda Mb^3\left[-\frac{6{\dot{b}}^2}{M^2b^2}-\frac{2\lambda}{\kappa}+\frac{1}{\kappa}\left(X^2+3Y^2-2XY^3\right)\right]+MXb^3Y^3\left(\frac{\dot{\phi}^2}{M^2X^2}-2V\left(\phi\right)\right)\,.
\label{LA2}
\end{equation}
After the Legendre transformation, the total Hamiltonian can be obtained as follows
\begin{align}
\mathcal{H}_T=&-\frac{M}{24\lambda b}p_b^2+\frac{MX}{4b^3Y^3}p_{\phi}^2-\frac{\lambda Mb^3}{\kappa}\left(X^2+3Y^2-2XY^3-2\lambda\right)+2MXb^3Y^3V\left(\phi\right)\nonumber\\
&+\lambda_Mp_M+\lambda_Xp_X+\lambda_Yp_Y\,,
\label{HTscalar}
\end{align}
where $p_\phi$ is the conjugate momentum of the variable $\phi$. Note that $\lambda_M$, $\lambda_X$, and $\lambda_Y$ are the Lagrange multipliers associated with the primary constraints as in the case of a perfect fluid description.

The complete constraint analysis of this system has been carried out in Ref.~\cite{Albarran:2018mpg}. It turns out that the Hamiltonian is again a first class constraint as expected and it becomes a restriction in the Hilbert space at the quantum level. $p_M$ is another first class constraint and it corresponds to a gauge degree of freedom which can be fixed by assuming the lapse function $M$ to be a constant. It should be stressed that in the EiBI theory, there are two additional second class constraints which correspond to two algebraic relations in the theory. In the perfect fluid description, these second class constraints are given by Eqs.~\eqref{classicalconstr}. In the scalar field description, on the other hand, these constraints are also given by Eqs.~\eqref{classicalconstr} but one has to substitute the energy density and the pressure by their scalar field counterparts, $\rho_\phi$ and $p_\phi$, respectively. The explicit expressions of these second class constraints are given in Eqs. (3.9) and (3.10) in Ref.~\cite{Albarran:2018mpg}. Essentially, once we introduce the Dirac brackets to promote the phase space functions to quantum operators, the second class constraints can be regarded as zero operators \cite{Diraclecture}. As a result, the total Hamiltonian can be significantly simplified \cite{Albarran:2018mpg}
\begin{equation}
\mathcal{H}_T=M\left(-\frac{p_b^2}{24\lambda b}+\frac{X}{Y^3}\frac{p_\phi^2}{4b^3}+\frac{2\lambda b^3}{\kappa}\left(\lambda-Y^2\right)\right)+\lambda_Xp_X+\lambda_Yp_Y\,.
\label{HT2scalar}
\end{equation}
Note that at the quantum level, the last two terms can be omitted since $p_X$ and $p_Y$ are also second class constraints. We shall emphasize that Eq.~\eqref{HT2scalar} is still not the final expression of the WDW equation that we are going to study. The final expression of the WDW equation is expected to be a partial differential equation of $b$ and $\phi$. After inserting $p_X\sim0$ and $p_Y\sim0$, there remain two variables $X$ and $Y$ in Eq.~\eqref{HT2scalar}. Technically, we have to use again the second class constraints to relate these two variables to the phase space variables $b$, $p_b$, $\phi$, and $p_\phi$. It can be expected that these relations would depend on the cosmological models under consideration and also on the scalar field potential that we choose in the model. Different choices of the cosmological solutions and potentials certainly change the expressions of the second class constraints, hence change the quantization of the system and also the expression of the WDW equation. In the following two subsections, we will first consider the cosmological solution near the singularity of the auxiliary metric for a negative $\kappa$ and rewrite the WDW equation as a partial differential equation from which the wave function can be solved. A similar study for a positive $\kappa$ will be presented in the subsection \ref{subsec.scalarp}.

\subsection{The WDW equation for $\kappa<0$}
The energy density and its pressure can be described by a scalar field $\phi$ and its potential $V(\phi)$ via the following relations
\begin{align}
\rho_\phi&=\frac{\dot{\phi}^2}{2N^2}+V\left(\phi\right)\,,\label{rhophi}\\
p_\phi&=\frac{\dot{\phi}^2}{2N^2}-V\left(\phi\right)\,.\label{pphi}
\end{align}
Note that we have used a lower index $\phi$ to highlight that the energy density and the pressure are described through the dynamics of a scalar field. For the EiBI gravity with a negative $\kappa$, the physical Hubble rate near the bounce $a\rightarrow a_{m1}$ $(b\rightarrow0)$ is approximated as in Eq.~\eqref{bouncehubble}. On the other hand, the conservation of the energy-momentum tensor implies that the energy density and the pressure, if expressed as a function of the scale factor $a$, read $\rho_\phi\approx\rho_0a^{-3(1+w)}$ and $p_\phi\approx w\rho_0a^{-3(1+w)}$, respectively, where $\rho_0$ is an integration constant and $w$ stands for the equation of state defined by $w\equiv p_\phi/\rho_\phi$. Using the approximated Hubble rate Eq.~\eqref{bouncehubble} and Eqs.~\eqref{rhophi} and \eqref{pphi}, we obtain the asymptotic expression of the scalar field as a function of the scale factor $a$ near the bounce:
\begin{equation}
\phi\left(\delta a\right)=\phi_0+2A\sqrt{\delta a}\,,
\end{equation}
where $\delta a\equiv a-a_{m1}$ and
\begin{equation}
A\equiv\sqrt{\frac{3|\kappa|\rho_0\left(1+w\right)a_{m1}^{-3(1+w)-1}}{8}}=\sqrt{\frac{3\lambda\left(1+w\right)}{8a_{m1}}}\,.
\end{equation}
On the above equations, $\phi_0$ is an integration constant and it is the value of the scalar field when $\delta a=0$. Furthermore, using again Eqs.~\eqref{rhophi} and \eqref{pphi}, the scalar field potential can be expressed with the equation of state $w$ and it approaches a constant
\begin{equation}
V\left(\phi\right)\approx\frac{\lambda}{2|\kappa|}\left(1-w\right)\,,
\label{Vphi}
\end{equation}
when $\delta a\rightarrow 0$.  

Now, we shall rewrite the WDW equation \eqref{HT2scalar} in such a way that it only contains the phase space variables $b$, $\phi$, and their conjugate momenta. According to Eqs.~\eqref{8}, we get the following approximated equations:
\begin{equation}
\frac{X}{Y^3}\approx\frac{b^4}{\left(1+w\right)a_{m1}^4}\,,\qquad
Y^2\approx\frac{a_{m1}^2}{b^2}\,,\label{XYnegativek}
\end{equation}
when $b\rightarrow0$. Using Eq.~\eqref{XYnegativek}, the Hamiltonian can be written as
\begin{equation}
\mathcal{H}_T=M\left[-\frac{p_b^2}{24\lambda b}+\frac{b}{4\left(1+w\right)a_{m1}^4}p_\phi^2-\frac{2\lambda b^3}{|\kappa|}\left(\lambda-\frac{a_{m1}^2}{b^2}\right)\right]+\lambda_Xp_X+\lambda_Yp_Y\,.
\label{HT3scalarn}
\end{equation}
We would like to stress again that if we use the Dirac brackets to promote the phase space functions to quantum operators, the second class constraints can be treated as zero operators. That is why we can use Eqs.~\eqref{XYnegativek} to simplify the WDW equation. Actually, the second class constraints that correspond to the above substitutions are secondary constraints, which are also the equations of motion of the theory, i.e., Eqs.~\eqref{5}. More strictly speaking, one should first use the Dirac brackets to promote the phase space functions to quantum operators, obtaining the Hamiltonian operator $\hat{\mathcal{H}}_T$. Then one regards the second class constraints as zero quantum operators and make the substitutions mentioned above. In the aforementioned procedures, it looks like we are doing it the other way around, that is, making substitutions at the classical level then promoting the phase space functions to operators. In either case, the final expression of the WDW equation is the same. Therefore, we will use the Hamiltonian \eqref{HT3scalarn} to construct the WDW equation.

After promoting the phase space functions to quantum operators and simplifying the Hamiltonian operator with the second class constraints, we construct the WDW equation as follows
\begin{equation}
\langle b\phi|\frac{\hat{\mathcal{H}}_T}{b}|\psi\rangle=0\,,
\end{equation}
and choose the following factor ordering
\begin{align}
\left(\frac{{\hat{p}_b}}{b}\right)^2&=-\hbar^2\left(\frac{1}{b}\frac{\partial}{\partial b}\right)\left(\frac{1}{b}\frac{\partial}{\partial b}\right)=-\hbar^2\left(\frac{\partial}{\partial x}\right)\left(\frac{\partial}{\partial x}\right)\,,\nonumber\\
{\hat{p}_\phi}^2&=-\hbar^2\left(\frac{\partial}{\partial\phi}\right)\left(\frac{\partial}{\partial\phi}\right)\,.
\end{align}
Note that we have defined a new variable
\begin{equation}
x=\frac{b^2}{2}\,,
\end{equation}
to label the scale factor of the auxiliary metric. Finally, the WDW equation becomes
\begin{equation}
\left[\frac{\hbar^2}{24\lambda}\partial_x^2-\frac{\hbar^2}{4\left(1+w\right)a_{m1}^4}\partial_{\phi}^2-\frac{4\lambda x}{|\kappa|}\left(\lambda-\frac{a_{m1}^2}{2x}\right)\right]\psi(x,\phi)=0\,.
\label{WDWfin}
\end{equation}
Near the singularity ($x\rightarrow0$) of the auxiliary metric, the WDW equation \eqref{WDWfin} can be further approximated as
\begin{equation}
\left[\frac{\hbar^2}{24\lambda}\partial_x^2-\frac{\hbar^2}{4\left(1+w\right)a_{m1}^4}\partial_{\phi}^2+\frac{2\lambda}{|\kappa|}a_{m1}^2\right]\psi(x,\phi)=0\,.
\label{WDWfin2}
\end{equation}
In the next section, we will solve this equation \eqref{WDWfin2} to get the asymptotic behavior of the wave function $\psi(x,\phi)$ near the singularity ($x\rightarrow0$, $\phi\rightarrow\phi_0$) with a negative $\kappa$.

\subsection{The WDW equation for $\kappa>0$}\label{subsec.scalarp}
As shown explicitly in section \ref{class.sec}, the EiBI theory of gravity with a positive $\kappa$ resolves the big bang singularity quite differently as compared with the situation for a negative $\kappa$. When $\kappa>0$, the asymptotic behavior of the physical Hubble function is given by Eq.~\eqref{loiteringhubble} when the physical scale factor approaches its minimum value $a\rightarrow a_{m2}$ $(b\rightarrow0)$. Note that this happens at $t\rightarrow-\infty$ for a constant lapse function $N$. If the density and pressure of the fluid are related via a constant equation of state effectively, that is, $p_\phi=w\rho_\phi$, their relations with the physical scale factor can be obtained from the conservation equation: $\rho_\phi\approx\rho_0a^{-3(1+w)}$ and $p_\phi\approx w\rho_0a^{-3(1+w)}$. Combining Eqs.~\eqref{loiteringhubble}, \eqref{rhophi} and \eqref{pphi}, we obtain the asymptotic expression of the scalar field as a function of $\delta a\equiv a-a_{m2}$ as follows
\begin{equation}
\phi(\delta a)=\sqrt{B}\ln\delta a\,,
\end{equation}
where 
\begin{equation}
B\equiv\frac{3\kappa\rho_0\left(1+w\right)a_{m2}^{-3\left(1+w\right)}}{8}=\frac{3\lambda\left(1+w\right)}{8w}\,.
\end{equation}
It can be seen that $\phi\rightarrow-\infty$ when $\delta a$ and $b$ vanish. The scalar field potential approaches a constant when $a\rightarrow a_{m2}$ and its value depends on the equation of state $w$:
\begin{equation}
V(\phi)\approx\frac{\lambda}{2\kappa w}\left(1-w\right)\,.
\label{Vphi}
\end{equation}
Similar to what we have done in the previous subsection, we have to express $X/Y^3$ and $Y^2$ in Eq.~\eqref{HT2scalar} in terms of $b$, $\phi$, and their conjugate momenta. To do this, we use Eqs.~\eqref{12} to get
\begin{equation}
\frac{X}{Y^3}\approx\frac{1+w}{w}\,,\qquad Y^2\approx\frac{a_{m2}^2}{b^2}\,.\label{XYkn}
\end{equation}
Substituting \eqref{XYkn} into the Hamiltonian, we obtain
\begin{equation}
\mathcal{H}_T=M\left[-\frac{p_b^2}{24\lambda b}+\left(\frac{1+w}{4w}\right)\frac{p_\phi^2}{b^3}+\frac{2\lambda b^3}{\kappa}\left(\lambda-\frac{a_{m2}^2}{b^2}\right)\right]+\lambda_Xp_X+\lambda_Yp_Y\,.
\label{HT3scalar}
\end{equation}

To proceed, we use the following factor ordering:
\begin{equation}
\langle b\phi|b^3\hat{\mathcal{H}}_T|\psi\rangle=0\,,
\end{equation}
and
\begin{align}
b^2{\hat{p}_b}^2&=-\hbar^2\left(b\frac{\partial}{\partial b}\right)\left(b\frac{\partial}{\partial b}\right)=-\hbar^2\left(\frac{\partial}{\partial z}\right)\left(\frac{\partial}{\partial z}\right)\,,\\
{\hat{p}_\phi}^2&=-\hbar^2\left(\frac{\partial}{\partial\phi}\right)\left(\frac{\partial}{\partial\phi}\right)\,.
\end{align}
Note that we have defined a new variable
\begin{equation}
z=\ln b\,,
\end{equation}
and it can be seen that $z\rightarrow-\infty$ when $b\rightarrow0$. Finally, the WDW equation can be expressed as
\begin{equation}
\left[\frac{\hbar^2}{24\lambda}\partial_z^2-\frac{\left(1+w\right)\hbar^2}{4w}\partial_{\phi}^2-\frac{2\lambda a_{m2}^2e^{4z}}{\kappa}\right]\psi(z,\phi)=0\,.
\label{WDWfinp}
\end{equation}

\section{The wave functions in the perfect fluid description}\label{solvwdwp}
In the previous sections, we have obtained the asymptotic expressions of the WDW equation near the singularity ($b\rightarrow0$) of the auxiliary metric or equivalently the physical connection. For a negative $\kappa$, we have derived the WDW equations \eqref{wdwperfctflkn} and \eqref{WDWfin2}, by assuming that the matter field is governed by a perfect fluid and a scalar field, respectively. On the other hand, for a positive $\kappa$, the corresponding WDW equations with a perfect fluid and a scalar field have been obtained in Eqs.~\eqref{wdwpefkp} and \eqref{WDWfinp}, respectively. We will solve the wave functions for all these WDW equations and see whether the wave functions satisfy the DW boundary condition, i.e., the wave functions vanish, near the configuration of the singularity of the auxiliary metric. Let us first consider the cases in which the matter field is described by a perfect fluid and solve the WDW equations \eqref{wdwperfctflkn} and \eqref{wdwpefkp}.

\subsection{ The $\kappa<0$ case}

When the matter content is governed by a perfect fluid, the WDW equation for a negative $\kappa$ that we will take into account is given by Eq.~\eqref{wdwperfctflkn}. The general solution can be written as a linear combination of two independent solutions as follows
\begin{equation}\label{sol1}
\psi\left(y\right)=y^{\frac{1}{2}}\left[C_{1}J_{\frac{3}{8}}\left(3\sqrt{2}y^{\frac{4}{3}}\right)+C_{2}Y_{\frac{3}{8}}\left(3\sqrt{2}y^{\frac{4}{3}}\right)\right]\,,
\end{equation}
where $C_1$ and $C_2$ are constants. The functions $J_{\nu}\left[f\left(y\right)\right]$ and $Y_{\nu}\left[f\left(y\right)\right]$ are Bessel functions of first kind and second kind, respectively, with order $\nu=3/8$ and argument $f\left(y\right)=3\sqrt{2}y^{\frac{4}{3}}$. Near the singularity, i.e., $y\rightarrow0$, the two independent solutions can be approximated as follows \cite{Abramow}:
\begin{equation}
y^{\frac{1}{2}}J_{\frac{3}{8}}\left(3\sqrt{2}y^{\frac{4}{3}}\right)\approx\frac{3^{3/8}}{2^{3/16}\Gamma\left(\frac{11}{8}\right)}y\,,\qquad
y^{\frac{1}{2}}Y_{\frac{3}{8}}\left(3\sqrt{2}y^{\frac{4}{3}}\right)\approx-\frac{2^{3/16}\Gamma\left(\frac{3}{8}\right)}{3^{3/8}\pi}\,.\label{asympY1}
\end{equation}
It can be seen that when $y\rightarrow0$, the solution $\sqrt{y}\,J_{3/8}\left[f\left(y\right)\right]$ vanishes, while the other solution $\sqrt{y}\,Y_{3/8}\left[f\left(y\right)\right]$ approaches a non-zero constant. Therefore, the wave function \eqref{sol1} satisfies the DW condition near the singularity as long as one assumes $C_2=0$. 

\subsection{ The $\kappa>0$ case}

On the other hand, the WDW equation for a positive $\kappa$ within the perfect fluid description is given by Eq.~\eqref{wdwpefkp}. Depending on the value of the parameter $c_2$, which is positive (see Eq.~\eqref{c2definition}), the general solution can be categorized as follows 
\begin{align}
\psi\left(z\right)=z^{\frac{1}{2}}\left(\tilde{D}_{1}z^{\frac{1}{2}\sqrt{1-4c_{2}}}+\tilde{D}_{2}z^{-\frac{1}{2}\sqrt{1-4c_{2}}}\right)\,, \ \ \ \   \qquad &\textrm{if} \quad 0<c_2<\frac{1}{4}\,,\label{sol2pos} \\
\psi\left(z\right)=z^{\frac{1}{2}}\left(\bar{D}_{1}+\bar{D}_{2}\ln{z}\right), \qquad \qquad \qquad \qquad \qquad &\textrm{if} \qquad c_{2}=\frac{1}{4}\,,\label{sol3zero} \\ 
\psi\left(z\right)=z^{\frac{1}{2}}\left(D_{1}z^{\frac{i}{2}\sqrt{\left\vert1-4c_{2}\right\vert}}+D_{2}z^{-\frac{i}{2}\sqrt{\left\vert1-4c_{2}\right\vert}}\right), \qquad &\textrm{if} \qquad c_2>\frac{1}{4}\,, \label{sol2neg}
\end{align}
where $D_{1}$, $D_{2}$, $\tilde{D}_{1}$, $\tilde{D}_{2}$, $\bar{D}_1$, and $\bar{D}_2$ are integration constants. Next, we will consider different values of $c_2$ and investigate whether the wave function is able to satisfy the DW boundary condition near the singularity where $z\rightarrow0$. 
\begin{itemize}
\item If $0<c_{2}<1/4$, the general solution of the wave function is given by Eq.~\eqref{sol2pos}. It can be shown that for each independent solution, the variable $z$ has a positive power. Therefore, the general solution vanishes when $z\rightarrow0$, satisfying the DW condition at the singularity.
\item If $c_2=1/4$, the general solution is given by Eq.~\eqref{sol3zero} and the DW condition at $z\rightarrow0$ is unambiguously satisfied due to the factor $\sqrt{z}$.
\item If $c_2>1/4$, the general solution is given by Eq.~\eqref{sol2neg} and the power of $z$ is complex. In consequence, the wave function acquires an oscillating behavior described by the imaginary part of the power of $z$. However, the modulus of the wave function behaves as $|\psi|\approx \sqrt{z}$. Therefore, when $z\rightarrow0$, the modulus of the wave function vanishes and the DW condition is fulfilled.  
\end{itemize}
Consequently, in a perfect fluid description and when $\kappa>0$, the wave function always satisfies the DW condition at the singularity of the auxiliary metric.

\section{The wave functions in the scalar field description}\label{solvescalar}
For the scalar field description, the asymptotic expressions of the WDW equations near the singularity are given by Eqs.~\eqref{WDWfin2} and \eqref{WDWfinp}, corresponding to a negative and a positive value of $\kappa$, respectively. The WDW equations are partial differential equations with two independent variables. We will prove that even in these general cases in which one more degree of freedom is included into the system, the DW boundary condition can still be satisfied near the singularity of the auxiliary metric. 

\subsection{ The $\kappa<0$  case}

The WDW equation in the scalar field description with a negative $\kappa$ is given by the partial differential Eq.~\eqref{WDWfin2}. The general solution to Eq.~\eqref{WDWfin2} can be obtained by using the separation of variables such that the total wave function can be decomposed as a series of products of two single variable functions
\begin{equation}\label{sum3}
\psi\left(x,\phi\right)=\sum_{k}C_{k}\left(x\right)\varphi_{k}\left(\phi\right)\,,
\end{equation}
where $C_{k}\left(x\right)$ and $\varphi_{k}\left(\phi\right)$ are the solutions to the following two ordinary differential equations
\begin{equation}
\left(\frac{\hbar^{2}}{24\lambda}\partial_{x}^{2}+\frac{2\lambda}{\left\vert\kappa\right\vert}a_{m1}^2-k\right)C_{k}\left(x\right)=0\,,\qquad
\left[-\frac{\hbar^{2}}{4\left(1+w\right)a_{m1}^{4}}\partial_{\phi}^{2}+k\right]\varphi_{k}\left(\phi\right)=0\,,\label{matterpart3}
\end{equation}
and $k$ corresponds to the decoupling constant. The above ordinary differential equations \eqref{matterpart3} can be solved to get the solution of the gravitational part $C_{k}\left(x\right)$
\begin{equation}\label{solgrav3}
C_{k}\left(x\right)=\left\{E_{1,k}\exp\left[\sqrt{\frac{24\lambda}{\hbar^{2}}\left(k-\frac{2\lambda}{\left\vert\kappa\right\vert}a_{m1}^2\right)}x\right]+E_{2,k}\exp\left[-\sqrt{\frac{24\lambda}{\hbar^{2}}\left(k-\frac{2\lambda}{\left\vert\kappa\right\vert}a_{m1}^2\right)}x\right]\right\}\,,
\end{equation}
and the solution of the matter part $\varphi_{k}\left(\phi\right)$
\begin{equation}\label{solmatter3}
\varphi_{k}\left(\phi\right)=\left\{F_{1,k}\exp\left[\frac{2a_{m1}^2\sqrt{\left(1+w\right)k}}{\hbar}\phi\right]+F_{2,k}\exp\left[-\frac{2a_{m1}^2\sqrt{\left(1+w\right)k}}{\hbar}\phi\right]\right\}\,.
\end{equation}

It should be stressed that near the singularity of the auxiliary metric, the scalar field approaches a constant and the auxiliary scale factor vanishes, i.e., $\phi\rightarrow\phi_0$ and $x\rightarrow0$. Therefore, the matter part of the wave function $\varphi_k\left(\phi\right)$ is well-defined and the gravitational part tends to a constant value $E_{1,k}+E_{2,k}$. Since the total wave function is constructed by the product of $C_{k}$ and $\varphi_k$, the DW boundary condition can be satisfied as long as one requires $E_{1,k}=-E_{2,k}$. 

\subsection{ The $\kappa>0$ case}
For a positive value of $\kappa$, the asymptotic expression of the WDW equation in the scalar field description near the singularity of the auxiliary metric is given by Eq.~\eqref{WDWfinp}. Again, the partial differential equation can be solved by using the separation of variables. The total wave function can be decomposed as a series of products of the solutions corresponding to the gravitational part and matter part: 
\begin{equation}\label{sum4}
\psi\left(z,\phi\right)=\sum_{m}Q_{m}\left(z\right)\xi_{m}\left(\phi\right)\,,
\end{equation}
where $Q_{m}\left(z\right)$ and $\xi_{m}\left(\phi\right)$ are, respectively, the gravitational and matter part of the wave function. In this regard, the WDW equation \eqref{WDWfinp} can be decoupled into two ordinary differential equations as follows
\begin{equation}
\left(\frac{\hbar^{2}}{24\lambda}\partial_{z}^{2}-\frac{2\lambda a_{m2}^2e^{4z}}{\kappa}-m\right)Q_{m}\left(x\right)=0\,,\qquad
\left[-\frac{\left(1+w\right)\hbar^{2}}{4w}\partial_{\phi}^{2}+m\right]\xi_{m}\left(\phi\right)=0\,,\label{matterpart4}
\end{equation}
where $m$ is the value of the decoupling constant. The general solution to the gravitational part can be written in terms of the modified Bessel functions $I_{\mu}\left[g\left(z\right)\right]$ and $K_{\mu}\left[g\left(z\right)\right]$ as follows \cite{Abramow}:
\begin{equation}\label{solgrav4}
Q_{m}\left(z\right)=G_{1,m}I_{\mu}\left[g\left(z\right)\right]+G_{2,m}K_{\mu}\left[g\left(z\right)\right]\,,
\end{equation}
where $G_{1,m}$ and $G_{2,m}$ are integration constants. The order $\mu$ and the argument $g\left(z\right)$ of the modified Bessel functions can be explicitly expressed as 
\begin{align}
\mu=&\frac{\sqrt{6\lambda m}}{\hbar}\,,\label{mu4}\\
g\left(z\right)=&\sqrt{\frac{12\lambda^2 a_{m2}^2}{\hbar^2\kappa}}e^{2z}\,.\label{arg4}
\end{align}
On the other hand, the solution of the matter part can be solved as follows
\begin{equation}\label{solmatter4}
\xi_{m}\left(\phi\right)=\left\{H_{1,m}\exp\left[\frac{2}{\hbar}\sqrt\frac{wm}{\left(1+w\right)}\phi\right]+H_{2,m}\exp\left[-\frac{2}{\hbar}\sqrt\frac{wm}{\left(1+w\right)}\phi\right]\right\}\,,
\end{equation}
where $H_{1,m}$ and $H_{2,m}$ are integration constants. Note that the scalar field $\phi\rightarrow-\infty$ near the singularity of the auxiliary metric.

To further proceed, we assume that the decoupling constant $m$ is a real number. This assumption is fully physical as $m$ has dimension of energy. Under this assumption, the order $\mu$ acquires either a non-negative real value when $m\ge0$, or a purely imaginary value when $m<0$. Depending on the value of $m$, the asymptotic expressions of the modified Bessel functions at small arguments ($z\rightarrow-\infty$ and $g(z)\rightarrow0$) are given as follows \cite{Abramow}
\begin{align}
I_{\mu}\left[g\left(z\right)\right]\approx\frac{1}{\Gamma\left(\mu+1\right)}\left[\frac{g\left(z\right)}{2}\right]^{\mu}\qquad&\textrm{when $\mu\neq-1,-2,-3,...$}\,,\label{asympI4} \\
K_{\mu}\left[g\left(z\right)\right]\approx\frac{\Gamma\left(\mu\right)}{2}\left[\frac{g\left(z\right)}{2}\right]^{-\mu} \qquad&\textrm{when $\mu$ is real and positive}\,, \label{asympK1}\\
K_{\mu}\left[g\left(z\right)\right]\approx-\ln{\left[g\left(z\right)\right]}\qquad & \textrm{when $\mu=0$}\,,\label{asympK0}\\
K_{\mu}\left[g\left(z\right)\right]\approx-\left[\frac{\pi}{\nu\sinh{(\pi\nu)}}\right]^{\frac{1}{2}}\sin\left[\nu\ln{\left(\frac{g(z)}{2}\right)}\right]\qquad&\textrm{when $\mu=i\nu$ is purely imaginary}\,, \label{asympK1im}
\end{align}
where $\Gamma\left(\alpha\right)$ stands for the Gamma function. In the following, we will investigate the behaviors of the total wave function for different values of the decoupling constant $m$.

\begin{itemize}
\item  If $m<0$, the matter part of the wave function \eqref{solmatter4} turns out to be a plane wave solution whose oscillating amplitude is constant. As for the gravitational part, the order $\mu$ becomes imaginary, and therefore according to Eqs.~\eqref{asympI4} and \eqref{asympK1im}, the modified Bessel functions $I_{\mu}\left[g\left(z\right)\right]$ and $K_{\mu}\left[g\left(z\right)\right]$ are both rapidly oscillating functions with a non-zero constant modulus. In consequence, for $m<0$ the total wave function does not vanish at the singularity and the DW condition cannot be satisfied.
\item If $m=0$, the solution to the matter part is given by 
\begin{equation}\label{solmatter4mequal0}
\xi_{0}\left(\phi\right)=\tilde{H}_{1,0}+\tilde{H}_{2,0}\phi\,,
\end{equation}
where $\tilde{H}_{1,0}$ and $\tilde{H}_{2,0}$ are integration constants. On the other hand, the asymptotic expressions of the modified Bessel functions with a zero order are given by Eqs.~\eqref{asympI4} and \eqref{asympK0}. Obviously, as $z\rightarrow-\infty$ and $\phi\rightarrow-\infty$, we get $I_0\left(0\right)\rightarrow1$, $K_0\left(0\right)\rightarrow\infty$, and $\xi_{0}\left(-\infty\right)\rightarrow\pm\infty$. Therefore, the DW condition cannot be satisfied.

\item If $m>0$, the matter part of the wave function turns out to be exponential functions. If we assume $H_{2,m}=0$, the growing part of the solution when $\phi\rightarrow-\infty$ is removed. On the other hand, the order $\mu$ of the modified Bessel functions in the gravitational part is a positive and real number. In this case, it can be seen from Eqs.~\eqref{asympI4} and \eqref{asympK1} that the modified Bessel function $I_{\mu}\left[g\left(z\right)\right]$ vanishes when $g(z)\rightarrow0$, while $K_{\mu}\left[g\left(z\right)\right]$ diverges. Consequently, one has to further choose $G_{2,m}=0$ in order to ensure the DW condition near the singularity of the auxiliary metric.
\end{itemize}

In summary, we have found that if $m\le0$, it is impossible to obtain a wave function satisfying the DW boundary condition at the singularity of the auxiliary metric. In fact, one is supposed to impose an additional condition on the decoupling constant, i.e., $m>0$, such that the DW condition is able to be satisfied.

\section{Conclusions}\label{conclusionsec}
In the context of the EiBI gravity, it has been shown that the propagation of gravitational waves would be affected by the geometry of the auxiliary metric, which is compatible with the affine connection of the theory. Therefore, even though the big bang singularity can be resolved, the singularity is present in the auxiliary metric and it has an important consequence on the behavior of the linear perturbations. The linear perturbations, including the tensor modes, turn out to be unstable in the non-singular solutions within the EiBI theory. In this paper, we consider the quantum geometrodynamical approach in the context of the EiBI gravity. Note that the Born-Infeld type of theories seem to have intrinsic Noether symmetries as shown recently in \cite{Capozziello:2014qla}. This also supports the choice of the EiBI action in this paper. Our motivation is to see whether or not the singularity in the auxiliary metric can be ameliorated by the quantum effects. It turns out the answer is yes and therefore, the linear instabilities of the physical metric, which are associated with the singular behavior of the auxiliary metric, would be resolved by the same token. 

For the sake of completeness, we have considered two descriptions regarding the matter sector of the theory. In the perfect fluid description, the matter field is governed by a perfect fluid with a positive and constant equation of state. In the homogeneous and isotropic universe, the system is characterized by a single variable $b$, the scale factor of the auxiliary metric. In the second description, that is, the scalar field description, we introduce a scalar field degree of freedom to incorporate the matter sector which, in the classical level, describes the evolution of the corresponding perfect fluid in the perfect fluid description. In this setup, the system contains two canonical degrees of freedom, the scale factor $b$ and the scalar field $\phi$, spanning a two dimensional configuration space. 

In the framework of quantum geometrodynamical approach, the building block is the WDW equation describing the quantum evolution of the universe as a whole. Essentially, we start with the alternative EiBI action in the Einstein frame and derive the classical Hamiltonian for both descriptions mentioned above. The Hamiltonian constraint, which is a first class constraint, is regarded as a restriction on the Hilbert space and the WDW equation is derived by promoting all phase space functions to quantum operators. The commutation relations are constructed by using the Dirac brackets which are necessary for a system containing second class constraints. We have derived the asymptotic expressions of the WDW equations for the two matter descriptions, and for positive and negative values of the Born-Infeld parameter $\kappa$. For a negative value of $\kappa$, the physical metric bounces in the past at the classical level. The asymptotic expressions of the WDW equations near the bounce are given by Eqs.~\eqref{wdwperfctflkn} and \eqref{WDWfin2}, for the perfect fluid and the scalar field descriptions, respectively. For a positive value of $\kappa$, the physical metric acquires its minimum scale factor in the asymptotic past (the loitering effect). The approximated WDW equations are given by Eqs.~\eqref{wdwpefkp} and \eqref{WDWfinp}, for the perfect fluid and the scalar field descriptions, respectively. 

After deriving the WDW equations, we have studied the quantum behavior of the universe by solving the wave function as a solution to the WDW equations. We have found that for each WDW equation under consideration, wave functions which satisfy the DW boundary conditions at the singularity of the auxiliary metric can always be obtained. Therefore, the hidden singularity in the auxiliary metric is expected to be avoided at the quantum level and the linear instabilities are not harmful in the quantum world.

\acknowledgments
 
The work of IA was supported by a Santander-Totta fellowship ``Bolsas de Investiga\c{c}\~ao Faculdade de Ci\^encias (UBI) - Santander Totta". The work of MBL is supported by the Basque Foundation of Science Ikerbasque. She also would like to acknowledge the partial support from the Basque government Grant No. IT956-16 (Spain) and from the project FIS2017-85076-P (MINECO/AEI/FEDER, UE). CYC and PC are supported by Ministry of Science and Technology (MOST), Taiwan, through No. 107-2119-M-002-005, Leung Center for Cosmology and Particle Astrophysics (LeCosPA) of National Taiwan University, and Taiwan National Center for Theoretical Sciences (NCTS). CYC is also supported by MOST, Taiwan through No. 108-2811-M-002-682. PC is in addition supported by US Department of Energy under Contract No. DE-AC03-76SF00515. This paper is based upon work from COST action CA15117 (CANTATA), supported by COST (European Cooperation in Science and Technology).

\end{document}